\documentclass[aps,prb,twocolumn,groupedaddress]{revtex4-1}
\usepackage{graphicx}
\usepackage[colorlinks,linkcolor=red]{hyperref}

\begin{document}


\title{Large Intrinsic Valley Polarization and High Curie Temperature in Stable Two-dimensional Ferrovalley YX$_2$(X=I,Br and Cl)}


\author{Wen-Yu Liu}
\thanks{These authors contributed equally to this work}
\author{Bo Huang}
\thanks{These authors contributed equally to this work}
\author{Xu-Cai Wu}
\author{Shu-Zong Li}
\author{Hongxing Li}
\author{Zhixiong Yang}
\author{Wei-Bing Zhang}
\email{Corresponding author. zhangwb@csust.edu.cn}
\affiliation{Hunan Provincial Key Laboratory of Flexible Electronic Materials Genome Engineering, School of Physics and Electronic Sciences, Changsha University of Science and Technology, Changsha 410114, People's Republic of China.}


\date{\today}

\begin{abstract}
Ferrovalley materials with spontaneous valley polarization are crucial to valleytronic application. Based on first-principles calculations, we demonstrate that two-dimensional (2D) YX$_2$(X= I, Br,and Cl) in 2H structure constitute a series of  promising ferrovalley semiconductors with large spontaneous valley polarization  and high Curie temperature. Our calculations reveal that YX$_2$ are dynamically and thermally stable 2D ferromagnetic semiconductors with a Curie temperature above 200 K. Due to the natural noncentrosymmetric structure, intrinsic ferromagnetic order and strong spin orbital coupling, the large spontaneous valley polarizations of 108.98, 57.70 and 22.35 meV are also predicted in single-layer YX$_2$(X = I, Br and Cl),respectively. The anomalous valley Hall effect is also proposed based on the valley contrasting Berry curvature. Moreover, the ferromagnetism and valley polarization are found to be effectively tuning by applying a biaxial strain. Interestingly, the suppressed valley physics of YBr$_2$ and YCl$_2$ can be switched on via applying a moderate compression strain. The present findings promise YX$_2$ as competitive candidates for the further experimental studies and practical applications in valleytronics.
\end{abstract}


\maketitle

\section{\label{sec:intro}Introduction}
Valley, a local energy extreme in the conduction/valence band, is emerging as a new degree of freedom for next-generation electronic device.  Similar to the charge and spin of carriers, valley degree of freedom can be coded, stored and manipulated information, which is known as valleytronics\cite{NP_xu,valley_yao,CSR_yao}. The current works about valleytronics mainly focus on two-dimensional (2D) hexagonal materials such as graphene and 2H-phase transition metal dichalcogenides (TMDs) MX$_2$  due to their interesting valley-contrasting physics \cite{PRL_xiao,PRL_xiao2,PRB_yao,PRL_graphene}. Especially, MX$_2$ monolayers are the most promising candidates due to the spatial inversion-symmetry-broken and strong spin-orbit-coupling (SOC) effect. The unique structure of MX$_2$ monolayer leads to two degenerate but inequivalent valley states at K and K\_ in the momentum space. Due to the large separation of K/K\_, the valley states can effectively supress intervalley scatterings of phonon and impurity \cite{NP_xu,valley_yao,CSR_yao,PhysRevLett.110.016806}, which is benifit to design electronic devices with high storage density and low energy consumption.

However,  pristine  TMDs monolayers are intrinsically nonmagnetic, two degenerate valleys in hexagonal Brillouin zone are not polarized due to their time reversal symmetry. This limits their applications in direct information storage applications \cite{NP_xu,valley_yao,CSR_yao}. Clearly, the essential step in valleytronics is to break the degeneracy between the two valleys, that is, to achieve the valley polarization. Nowadays, many external strategies have been proposed to induce valley polarization. For example, optical pumping is extensively employed in experiments \cite{Zeng_NN,Feng_nc}. However, since optical pumping is a dynamic process, the induced carrier lifetime is very short, which is unsuitable for valleytronic applications. While valley polarization induced by an external magnetic field \cite{mag1,mag2,mag_3} is only 0.1$\sim$0.2 meV/T. More importantly, these attempts are volatility, which will limit device application seriously. Magnetic proximity and magnetic doping appeared as an alternative approach for nonvolatility application. However, magnetic doping \cite{mag_doping} tends to form clusters and increase scattering during carrier transport. The magnetic substrates for magnetic proximity effect \cite{proximity1,pro2} could enlarge the device size and increase energy consumption.

Recently, 2D ferrovalley materials with spontaneous valley polarization were proposed \cite{fv_duan}, which provides new opportunities to overcome the shortages mentioned above. The ferrovalley materials with spatial-inversion-symmetry-breaking and long-range ferromagnetism are expected to produce spontaneous valley polarization under SOC effect. Although some potential ferrovalley materials such as VSe$_2$\cite{fv_duan,VSe2_meng}, LaBr$_2$\cite{LaBr2},VSi$_2$N$_4$ \cite{PhysRevB.103.085421}, GdI$_2$ \cite{GdI2}, Cr$_2$Se$_3$ \cite{PhysRevB.104.075105},NdX$_2$(X=Se,S)\cite{NbX2},and MBr$_2$ (M=Ru,Os) \cite{yang_prb} have been reported, the candidates of ferrovalley materials are still scarce at present. Moreover, most of available materials possess small valley polarization and low Curie temperature. In addition, the prominent K/K\_ valleys in some ferrovalley materials do not locate at the valence band maximum (VBM) of energy band. These shortages limit the practical device application of ferrovalley materials. It is thus desirable to explore novel 2D ferrovalley materials with large valley polarization and high Curie temperature.

In the present work, we predict that single-layer(SL) 2H-YX$_2$(X = I, Br, and Cl) as a series of potential 2D ferrovalley semiconductors. The phonon dispersion calculations  and \emph{ab initio} molecular dynamics (MD) simulations are also performed to evaluate the stability of SL YX$_2$. The Curie temperature is also predicted by using Monte Carlo (MC) simulations and spin Hamiltonian. The valley polarizations of three SL YX$_2$(X = I, Br, Cl) are calculated and the underlying mechanism is revealed.  A spontaneous valley polarization of 109 meV  and a Curie temperature of 220K  are found in case of YI$_2$. Moreover, Berry curvature is also calculated and the anomalous valley Hall effect is proposed. Finally,we also explore the effect of biaxial strain on valley splitting and Curie temperature. Our calculations show that single-layer 2H-YX$_2$ are stable 2D ferrovalley semiconductors with large spontaneous valley polarization and high Curie temperature. These make them promising materials for future valleytronic applications.

\section{\label{comp}Computational Details}
The present calculations have been performed using the Vienna \emph{ab initio} simulation package (VASP) code \citep{vasp_CMS,vasp_PRB} within projector augmented-wave (PAW) method \cite{vasp_PAW,vasppawprb}.  General gradient approximations (GGA) in the Perdew-Burke-Ernzerhof (PBE) implementation \cite{PBE} were employed as the exchange correlation functional. To better take into account the strong electronic correction of Y-d electrons,  a simple rotationally invariant DFT + U method \cite{ldau} with effective Hubbard U of 2 eV was used. Meanwhile, we also performed the hybrid functional HSE06 calculations \cite{hse06,hse062} for comparison. A plane-wave basis set with a energy cutoff of 500 eV was used in the calculation. A mesh of  9$\times$9$\times$1 \emph{k}-points generated by the scheme of Monkhorst-Pack \cite{VASP_MP} was used to sample the Brillouin zone. All the atomic coordinates were optimized  until  the maximum force on all atoms was less than $5\times10^{-2}$  eV/\AA. A large vacuum space of at least 15 \AA~ was included in the supercell to avoid interaction between images. The phonon calculations have been performed using the finite displacement approach, as implemented in the Phonopy code \cite{phonopy}, in which a 2$\times$2$\times$1 supercell were employed. Furthermore,  we have performed  \emph{ab initio} molecular dynamics simulations at 300 K in canonical ensemble using Nos\'e heat bath scheme to examine the thermal stability of SL YX$_2$.  In these calculations, a large supercell of (4$\times$4) was used to minimize the constraint of  periodic boundary condition. To calculate the Berry curvature and anomalous Hall conductivity, the maximally localized Wannier functions(MLWFs) implemented in the WANNIER90 package \cite{Pizzi2020,Mostofi2014} was employed.

\section{\label{results}Results and Discussion}
\subsection{\label{structure}Structure and Stability}

\begin{figure}
\centering
\includegraphics[width=0.45\textwidth]{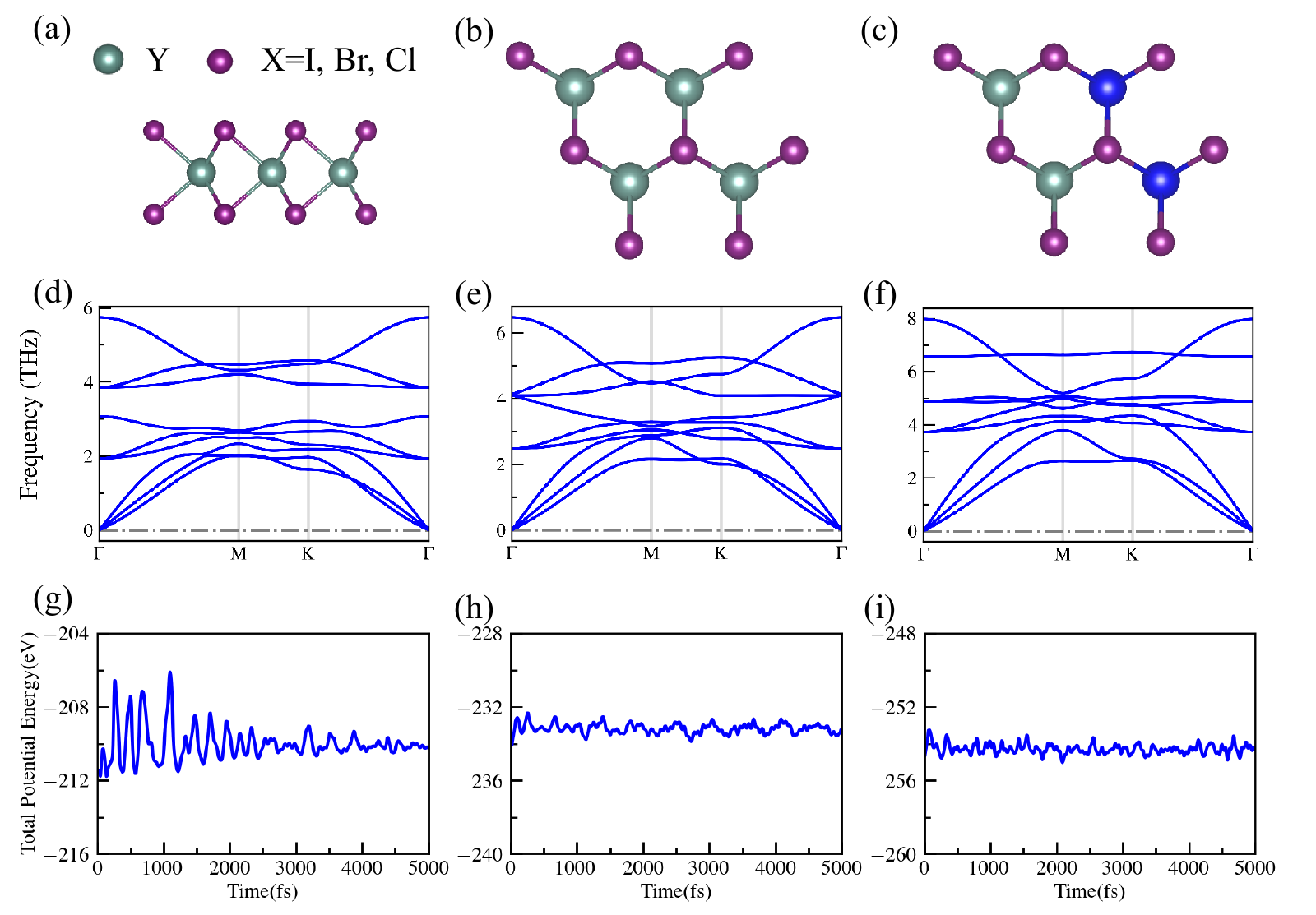}
\caption{\label{fig:stru}(Color online)Geometry and stability of single-layer(SL) YX$_2$.  The top (a) and side (b) view of structure of SL YX$_2$ . (c) The anti-ferromagnetic magnetic configuration in 2 $\times$2 supercell, which the atoms with different spin directions are labeled by different colors. Phonon spectrum of YI$_2$ , YBr$_2$ and  YCl$_2$ are given in (d),(e) and (f), respectively. Variations of the total potential energy of SL YX$_2$ with respect to simulation time during  \emph{ab initio} molecular dynamics  simulations are shown in (g),(h) and (i) .}
\end{figure}

As shown in Fig.~\ref{fig:stru}-(a) and -(b), SL 2H-YX$_2$(X=I, Br, Cl) possess  a hexagonal cell with space group P$\bar{6}$m2 (No.187), in which the inversion symmetry is absent. It contains one Y atom layer  and two X atom layers, where each Y atom is surrounded by six X atoms to form a trigonal prism. As given in Table.~\ref{table_data}, the lattice constants of SL YX$_2$ are 4.086, 3.949 and 3.853 \AA, respectively. The lattice constants of SL YX$_2$ increase with the element number of X, which can be attributed to the increasing atomic radius and the weakening reactivity  of X atoms.

\begin{table*}
\begin{ruledtabular}
\caption{\label{table_data}The calculated lattice constants, elastic constant, valley polarization ($\Delta_{val}$), magnetic exchange interaction (J), and Curie temperature for YX$_2$(X =I, Br, Cl)}
\begin{tabular}{ccccccccccc}
Material	&	$a_0$(\AA)	& $C_{11/22}$ (kBar) & $C_{12}$(kBar) & $C_{44}$(kBar)& $\Delta$E(meV)	&	$\Delta_{val}$ (meV)	&$J$	&$T_c$(K)\\
YI$_2$	&	4.086	&76.41 &38.94& 23.74&21.77	&	108.98	& 19.80	&230	\\
YBr$_2$	&	3.949&118.72 &43.87&37.43 &	19.57	&	57.70	& 18.68 &220	\\
YCl$_2$	&	3.853&127.25&54.40&36.43 &	19.38	&	22.35	& 18.07 &210	\\
\end{tabular}%
\end{ruledtabular}
\end{table*}%

To evaluate the stability of SL YX$_2$, we have performed the phonon dispersion, molecular dynamics and elastic constant calculations. As shown in Fig.~\ref{fig:stru}-(d) to Fig.~\ref{fig:stru}-(f), the imaginary frequency is found to be absent in the whole Brillouin Zone, indicating that three SL YX$_2$ are dynamically stable. Moreover,  room-temperature \emph{ab initio} molecular dynamics results also indicate total potential energies of SL YX$_2$ remain almost invariant during the simulation (Fig.~\ref{fig:stru}-(g) to Fig.~\ref{fig:stru}-(i)), which suggests that these SL YX$_2$ are thermally stable at 300K.  In addition, according to the Born criteria \cite{PhysRevLett.71.4182}, the elastic constants of  stable 2D materials should fulfill the conditions $C_{11}C_{22}-C_{12}^2>0$ and $C_{44}>0$. As shown in Table.~\ref{table_data}, these conditions are  satisfied for all three YX$_2$. Clearly, the dynamical, thermal and mechanical stability suggest that YX$_2$ can be realized experimentally even at room temperature.

\begin{figure}
\centering
\includegraphics[width=0.45\textwidth]{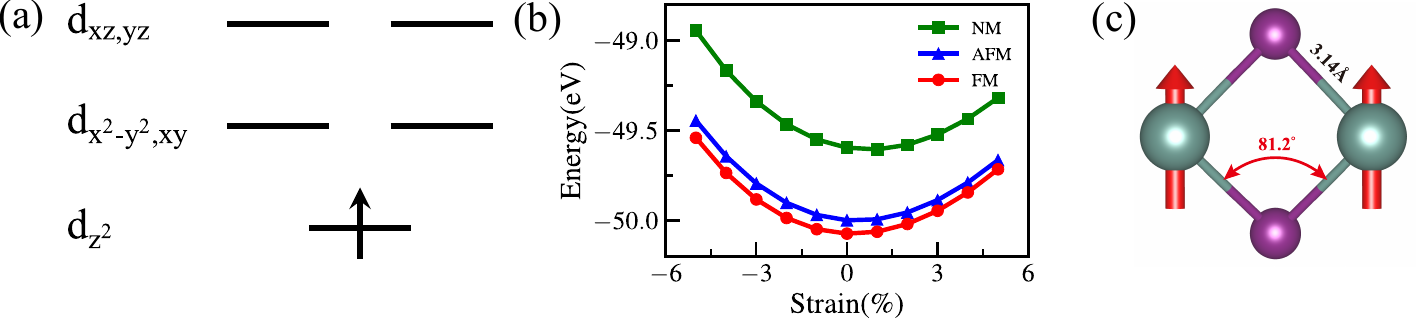}
\caption{\label{fig:mag}(Color online)(a) The splitting of d orbitals under the trigonal environment.(b) The change of total energy in different magnetic phases of YI$_2$ as a function of strain. (c) The paths for the mediation of nearest neighbour  magnetic exchange interactions in SL YI$_2$.   }
\end{figure}

\subsection{\label{mag}Magnetic Property}

According to the crystal field theory, $d$ orbital in trigonal environment will be split into A$_1$(d$_{z^2}$), E$_1$(d$_{xy}$, d$_{x^2-y^2}$ ), E$_2$(d$_{xz}$, d$_{yz}$). The electron configuration of an Y$^{2+}$ in YX$_2$ should be 4d$^1$ . Clearly, the only one electron will occupy the lowest energy d$_{z^2}$ orbital while the other orbitals are empty(Fig.~\ref{fig:mag}-(a)). This unique electron configuration will lead to a spin-polarized ground state , and the   magnetic moment of each Y atom is expected to be 1$\mu_B$ .

To confirm the magnetic ground state of SL YX$_2$, we considered  ferromagnetic (FM) and antiferromagnetic (AFM) state constructed in a 2 $\times$2 $\times$ 1 supercell as shown in Fig.~\ref{fig:stru}-(b) and (c).  The FM state is found to be the magnetic ground state of YX$_2$, following by the AFM state (Fig.~\ref{fig:mag}-(b)). The energies of FM state are 18.07(208.74), 18.68(174.51), 19.80(124.51) meV lower than those of AFM(nonmagnetic(NM) state) for YCl$_2$, YBr$_2$ and YI$_2$, respectively. Our calculations also predict a total magnetic moment of 1$\mu_B$ per YX$_2$ unit cell, which is in good agreement with the above analysis.

To reveal the underlying magnetic mechanism and also estimate the Curie temperature of YX$_2$, the Ising model and standard Metropolis Monte Carlo simulation are used. The nearest neighbour (NN) exchange interactions are extracted by fitting the total energies from DFT calculations for FM and AFM state to the spin Hamiltonian
\begin{equation}
\label{eq1}{
H=- J \sum \limits_{\langle ij\rangle} \vec{S_i}\cdot\vec{S_j}
}
\end{equation}
where \emph{J} is the exchange coupling constant and spin magnetic quantum number $\vec{S}$=1/2.

The magnetic energy  of four YX$_2$ formulas can be explicitly expressed as:
\begin{equation}
E_{FM}=E_0-12J\vec{S}^2
\end{equation}
\begin{equation}
E_{AFM}=E_0+4J\vec{S}^2
\end{equation}

\begin{figure}
\centering
\includegraphics[width=0.45\textwidth]{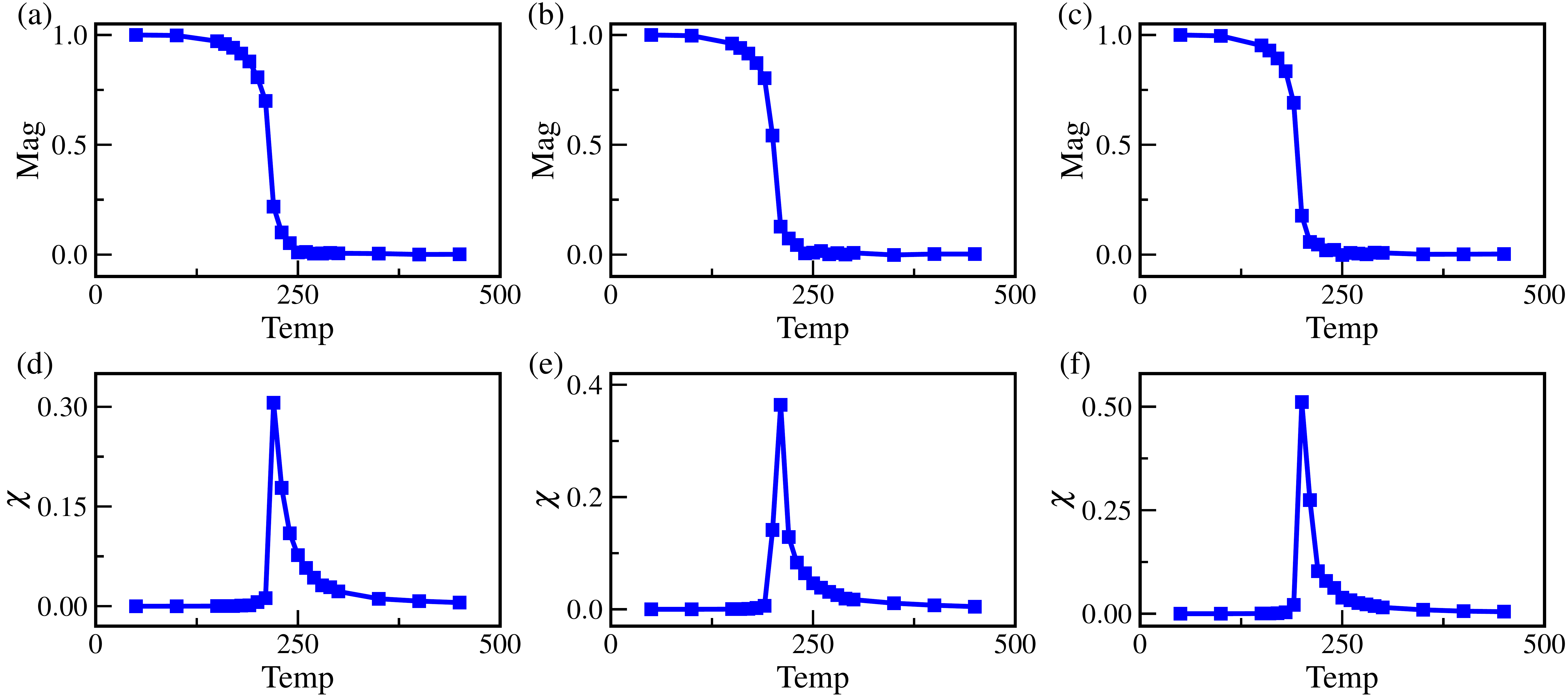}
\caption{\label{fig:mc}(Color online) The change of magnetic moment and magnetic susceptibility as a functional of temperature. The (a), (b) and (c) correspond to variance of magnetic moment of YI$_2$,YBr$_2$ and YCl$_2$, respectively, while the  (d),(e) and (f) correspond to the change of magnetic susceptibility. }
\end{figure}

As listed in Table \ref{table_data},  the $NN$ exchange interaction $J$  is predicted to be ferromagnetic. As shown in Fig.~\ref{fig:mag}-(c), the Y–X–Y (X=I,Br,Cl) bonding angles are 81, 84 and 86$^\circ$. Based on the Goodenough-Kanamori-Anderson (GKA) rules \cite{good},  ferromagnetic superexchange interaction with near 90$^\circ$ would dominate the interaction between Y atoms , thereby leading to the FM coupling. Using these DFT-derived magnetic exchange parameters, the Curie temperature is then estimated using a 40 $\times$40 $\times$ 1 supercell with periodic boundary conditions. For each temperature studied, the MC simulation involves for 10$^5$ MC steps per site to attain thermal equilibrium. The temperature-dependent mean magnetic moment and magnetic susceptibility are then obtained.  The temperature at which the mean magnetic moment drops drastically to nearly zero and the magnetic susceptibility peaks sharply is identified as  the Curie temperature. As shown in Fig.~\ref{fig:mc},   the Curie temperature(Tc) of SL YI$_2$, YBr$_2$ and YCl$_2$ are about 220, 210 and 200 K,  respectively. These values are much larger than that of most of available 2D ferromagnetic materials. For example, the T$_c$ of the recently  experimental discovered 2D ferromagnetic materials CrI$_3$ \cite{zhang_CrI3,exp_crI3}and Cr$_2$Ge$_2$Te$_6$ \cite{bilayer_fe}  is only 45 and 30K.  Therefore, the predicted high Curie temperature of SL YX$_2$ is quite benefit to the practical application in valleytronics and spintronics.

\subsection{\label{valley}Electronic Structure and Valley Splitting}

\begin{figure}
\centering
\includegraphics[width=0.45\textwidth]{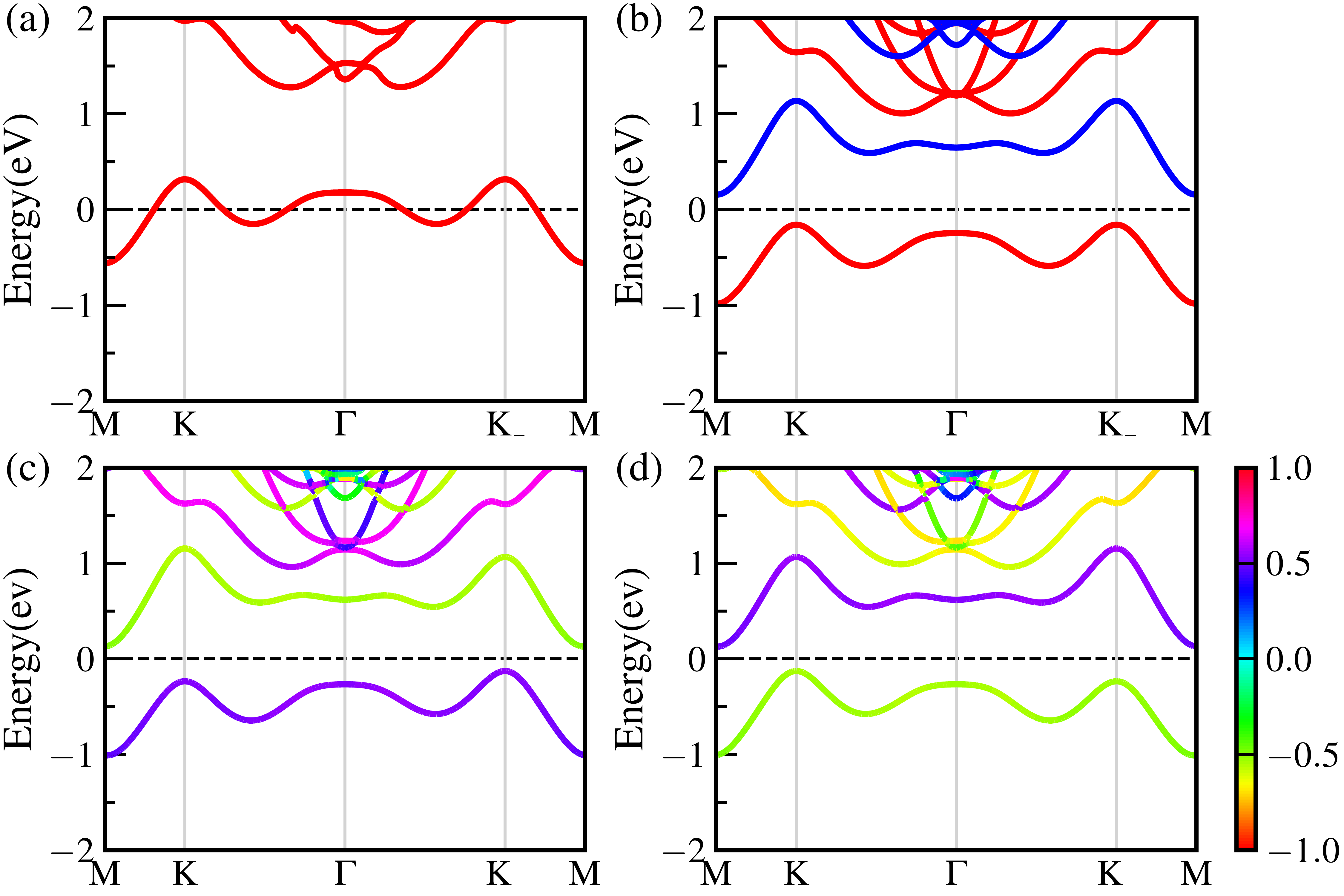}
\caption{\label{fig:band}(Color online)The band structure of SL YI$_2$. (a) without spin-polarization and SOC. (b) with spin-polarization but without SOC.Red and blue lines respresent the spin-up and  spin-down bands. (c) both with spin-polarization and  SOC. (d) is the same as (c) but with the opposite magnetization orientation. The Fermi level is set to 0 eV. }
\end{figure}

Now, turn our attention to the electronic structure. Since the band structures of YX$_2$ are very similar, we discuss YI$_2$ as an illustrative example. As shown in Fig.~\ref{fig:band}-(a), when neither spin polarization nor SOC is considered, a typical metallic state with a spin degenerate is found in YI$_2$. Energies at K/K\_ valleys are degenerate. As shown in Fig.~\ref{fig:band}-(b),when spin polarization is considered in the calculations,  a sizeable energy gap is formed and a clear exchange splitting can be found.  The K and K\_ valley still keep degenerate and have the same spin component.  The spin splitting $\Delta_{spin}$ of YI$_2$ , which is defined as the energy difference between  two spin directions in the K or K\_ valleys, are 1.30 eV. We can also find that the top-most valence band and the lowest conduction band are contributed by the spin-up and spin-down band, respectively. The  valence band maximum (VBM) is located at K or K\_, while the  conduction band minimum (CBM) is located at M. Therefore, SL YI$_2$ is a bipolar ferromagnetic semiconductor.

\begin{figure}
\centering
\includegraphics[width=0.45\textwidth]{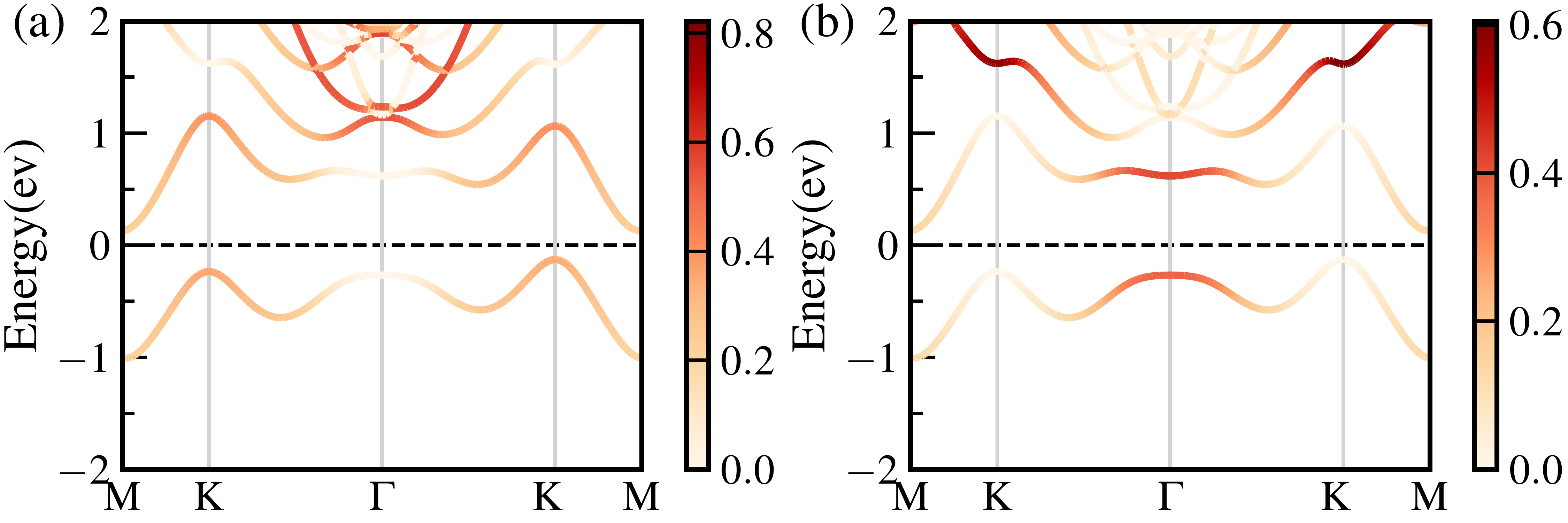}
\caption{\label{fig:fat_band}(Color online) The 4d orbital-resolved band structure of SL YI$_2$. (a) and (b) represent d$_{xy}$(d$_{x^2-y^2}$) and  d$_{z^2}$, respectively.}
\end{figure}

With spin polarization and SOC, the indirect semiconducting properties of SL YX$_2$ can be still found. The energy gap is about 0.255 eV.  We also confirm the electronic structure of SL YI$_2$ by using more accurate hybrid functional HSE06. The band structure is similar but the gap increases to 0.711 eV. The predicted  indirect band gaps of SL YX$_2$(X=Br,Cl) are 0.530 and 0.626 eV, respectively. Interestingly,  the valley degeneracy between the K and K\_ points is lifted as shown in Fig.~\ref{fig:band}-(c).   A valley splitting occurred spontaneously in SL YX$_2$ without any external tuning . Valley polarization $\Delta_{val}$ defined as the energy difference between K and K\_ valleys of YI$_2$, YBr$_2$, and YCl$_2$ were predicted to be 108.98, 57.75, and 22.35 meV, respectively. The valley polarization of YI$_2$ calculated by HSE06 functional is as large as 122 meV. This value is much larger than those of available ferrovalley materials such as VSe$_2$(78.2meV)\cite{VSe2_meng}, LaBr$_2$(33meV)\cite{LaBr2}, VSi$_2$N$_4$ (71 meV) \cite{PhysRevB.103.085421},and Cr$_2$Se$_3$(18.7 meV), comparable  with the values of GdI$_2$ (149 meV), and NdSe$_2$ (219 meV)\cite{NbX2}. It should be pointed out that the prominent valley in a qualified ferrovalley material should locate at the VBM/CBM.  As shown in Fig.~\ref{fig:Brband} and Fig.~\ref{fig:Clband}, the energies of $\Gamma$ of YBr$_2$ and YCl$_2$ are higher than that of K/K\_. This indicates that the prominent K/K\_valley is hidden, and YBr$_2$ and YCl$_2$ may be not suitable for practical valleytronics directly. When reversing magnetization orientation, the polarization states of K valley and K\_ valley also reverse. The energy of K valley is higher than that of K\_ valley, and the value of valley splitting remains unchanged as shown in Fig.~\ref{fig:band}-(d). Moreover, we can find that the  valence and conduction band is only contributed by one spin, which is also beneficial to detection and manipulation of valley. As illustrated in Fig. ~\ref{fig:fat_band}, Fig.~\ref{fig:Brfat_band} and Fig.~\ref{fig:Clfat_band}, both  valleys  at K and K\_ are primarily contributed by d$_{xy}$ and d$_{x^2-y^2}$ orbitals, while the electronic states  at  $\Gamma$ points are mainly contributed by d$_{z^2}$ orbitals.

The large spontaneous valley polarization in YX$_2$ can be sourced from the large magnetic exchange interaction and strong SOC effect. Without spin polarization and SOC, the single electron will occupy the twofold spin degenerate d$_{z^2}$, which forms a metallic state. With spin polarization but excluding SOC, a huge spin  splitting  up to 1.30 eV  induced by the magnetic exchange interaction occurs. However, the energetic degeneracy between K and K\_ is preserved since the E$_{\uparrow}$(K)=E$_{\uparrow}$(K\_),E$_{\downarrow}$(K)=E$_{\downarrow}$(K\_) in the absence of SOC. When including SOC but excluding spin polarization, the spatial inversion symmetry is broken and SOC still induces inequivalent valley at K and  K\_ . But two valleys are energetically degenerate with opposite spins due to the time reversal symmetry ,i.e., E$_{\uparrow}$(K)=E$_{\downarrow}$(K\_),E$_{\downarrow}$(K)=E$_{\uparrow}$(K\_).
With spin polarization and SOC, the time inversion symmetry of SL YI$_2$ is broken, and the energy of K and K\_ valley is no longer degeneracy.  The valley polarization occurs spontaneous without any external tuning. In the case of YI$_2$, the spin splitting caused by magnetic exchange interaction is much larger than that caused by SOC, which leads to the clean spin contribution in both the valence band and conduction band. Moreover, according to the analysis of orbital-resolved band,  the remarkable  spontaneous valley polarization in YI$_2$ can be understood by the relatively strong SOC effect within Y-d$_{xy}$ and d$_{x^2-y^2}$ orbitals combined with the magnetic exchange interaction of Y-d electrons.

\subsection{\label{berry}Berry curvature and valley anomalous Hall effect }

\begin{figure}
\centering
\includegraphics[width=0.45\textwidth]{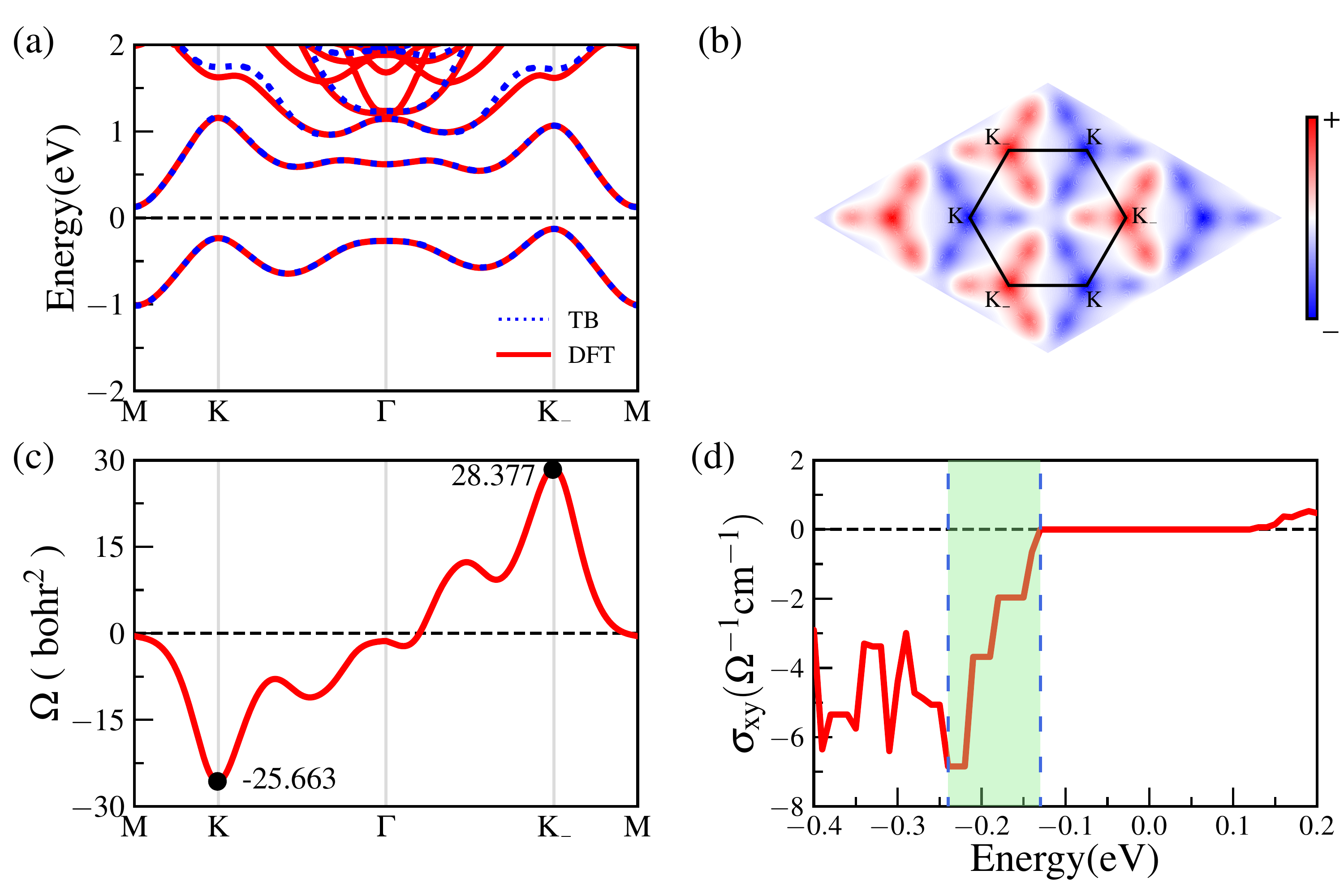}
\caption{\label{fig:berry}(Color online) The comparison of energy band calculated by PBE+U and Wannier fitting methods(a). The contour map  for the Berry curvatures in the Brillouin zone of the SL YI$_2$, and the corresponding Berry curvature along points of high symmetry lines is shown in (b) and (c). Anomalous Hall conductivity as a function of energy is present in (d).}
\end{figure}

To reveal the valley contrasting physics in 2H-YI$_2$, we calculated the Berry curvature , which is defined as \cite{PhysRevLett.49.405}
\begin{equation}\label{eq:kubo}
  \Omega(k)=-\sum_{n}\sum_{n\neq n^{\prime}}f_n \frac{2Im\langle\varphi_{nk}|\upsilon_x|\varphi_{n^{\prime}k}\rangle \langle\varphi_{n^{\prime}k}|\upsilon_y|\varphi_{nk}\rangle }{(E_n-E_{n^{\prime}})^2}
\end{equation}
where f$_n$ is the Fermi-Dirac distribution function,  $\varphi_{nk}$ is the periodic part of the Bloch wave function with eigenvalue E$_n$, $\upsilon_{x/y}$ is the velocity operator along the x/y direction. In the present work, the Berry curvatures of SL YI$_2$ was evaluated using the maximally localized Wannier function method. We first compared  the electronic band obtained in direct DFT calculation and Wannier fitting methods. As shown in Fig.~\ref{fig:berry}-(a), both results  matched well with each other near the Fermi level, which indicates the accuracy of the present calculation.

Berry curvature distributions in the 2D Brillouin region and along high symmetry points are shown in Fig.~\ref{fig:berry}-(b) and -(c). We can find that the Berry curvatures of K and K\_ valley possess opposite signs and the absolute values are 25.663 and 28.377 Bohr$^2$, respectively.  The non-degenerate and sizeable Berry curvature of K/K\_ valley with opposite signs leads to interesting valley contrasting physics of SL YI$_2$.

Due to the large valley polarization in SL YI$_2$, the Fermi level can shift between the energies of the valence band at the K and K\_ valleys with proper hole doping. Under an in-plane electric field, the holes at the K\_ valley will acquire an anomalous velocity proportional to the Berry curvature \cite{RevModPhys.82.1959}, i.e., $\upsilon_{\bot}\sim E \times \Omega(k)$.  The spin-down holes at the  K\_ valley will flow to the right edge of the sample.  The accumulated holes on the boundary generate a charge Hall current that can be detected by a positive voltage. Otherwise, when reversing the magnetization orientation , in the presence of an in-plane electric field, the spin-up holes at the K valley will flow to and be accumulated at the left edge of the sample. And a negative voltage can be measured.  The anomalous valley Hall effect thus can be realized in SL YI$_2$. Accordingly, the valley pseudospin can be detected and  manipulated selectively by electric measurement, which will provide a  basis for the application of the valleytronics, such as data storage.

By integrating berry curvature over the Brillouin zone
\begin{equation}\label{eq:avhc}
  \sigma_{xy}=\frac{e^2}{\hbar}\int_{BZ}\frac{d^2k}{(2\pi)^2}f(k)\Omega(k)
\end{equation}
 ,  one can obtain the anomalous valley Hall conductivity. As shown in Fig.~\ref{fig:berry}-(d), a fully spin- and valley-polarized Hall conductivity is generated in SL YI$_2$.

\subsection{\label{strain}Strain}

\begin{figure}
\centering
\includegraphics[width=0.5\textwidth]{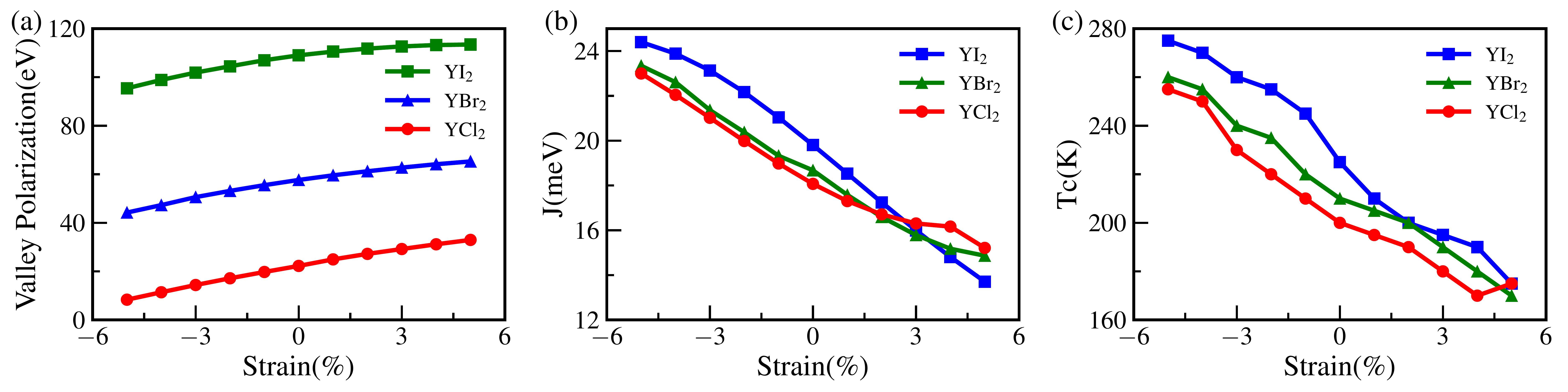}
\caption{\label{fig:strain}(Color online) The valley polarization (a), magnetic exchange constants (b) and Curie temperature (c) of YI$_2$ as a function of strains.}
\end{figure}

\begin{figure}
\centering
\includegraphics[width=0.45\textwidth]{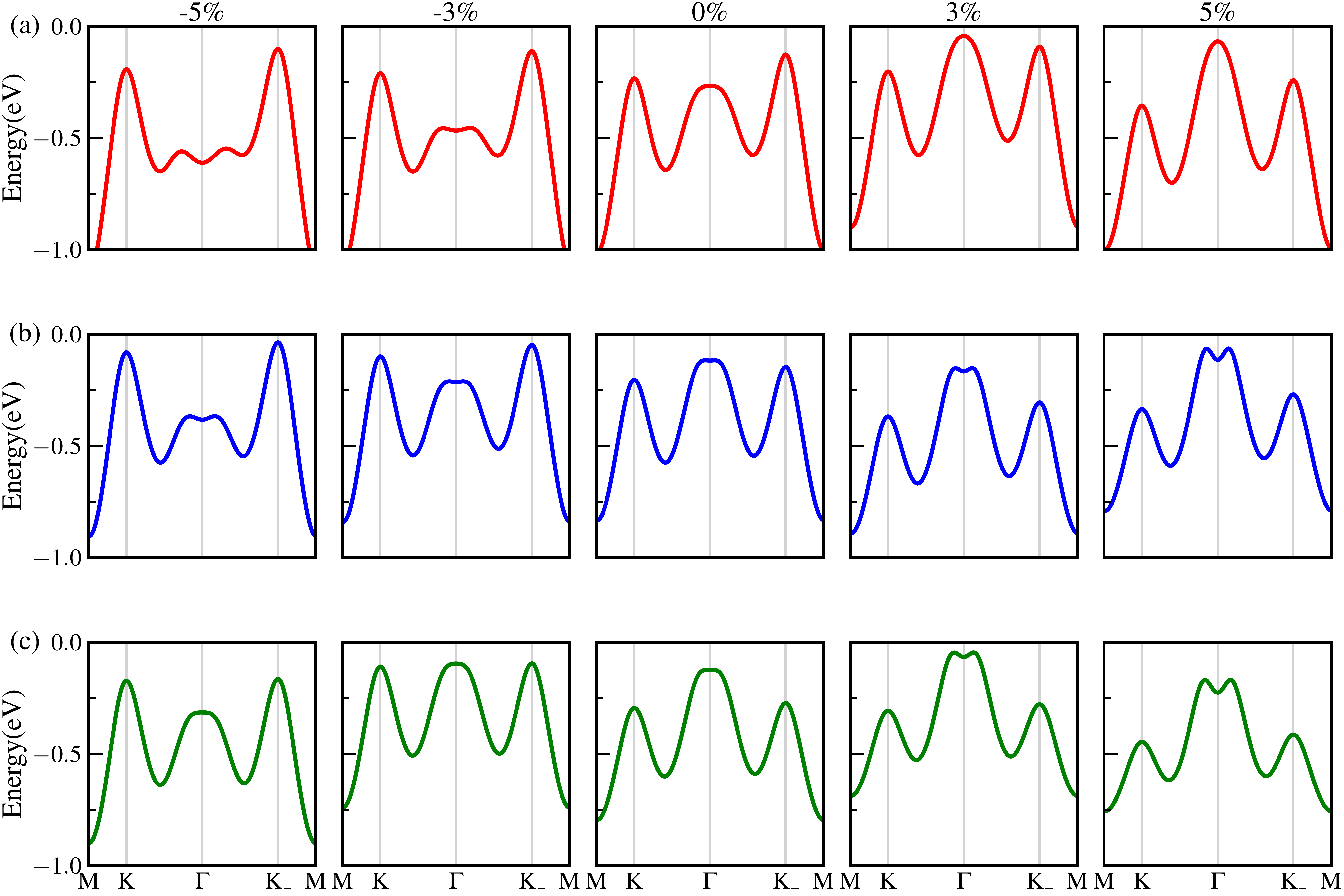}
\caption{\label{fig:band_strain}(Color online) The variance of band structure of YX$_2$ as a functional of strains. (a),(b) and (c) corresponds the results of YI$_2$, YBr$_2$ and YCl$_2$.}
\end{figure}
Furthermore, we also investigate the ferromagnetism and valley polarization of YX$_2$ under a biaxial strain. Fig.~\ref{fig:mag}-(b) gives the variation of total energy of YX$_2$ in FM, AFM and NM state  with different strains. Under a strain from -5\% to 5\% , the FM is always the lowest-energy state while the energy of NM state is  highest. This indicates that the ground state of YX$_2$ is FM state , which is robust to strain.

As shown in Fig.~\ref{fig:strain}-(a),under the strain of -5\%$\sim$5\%, the valley polarization  of SL YI$_2$ increases gradually from 91.56 to 113.22 meV. While the result of YBr$_2$ (YCl$_2$) can also be increase from 44.23(8.35) to 65.282(32.9) meV.  Nevertheless, we can also find that $\Gamma$ point gradually moves upwards VBM with strain and the VBM will change  from K/K\_ to $\Gamma$. For YI$_2$, the energy of $\Gamma$ is still lower than K and K\_valleys in the range of -5\%$\sim$3\% but become larger with the tensile  strain above 3\%. For YBr$_2$ and YCl$_2$, the corresponding threshold value of VBM transition is -1\% and -3\%. When compress strain are larger than the value, the K/K\_valleys become the VBM and the valley physics of YBr$_2$ and YCl$_2$ will be accessible.

We also evaluate the effect of strain on Curie temperature. As shown in  Fig.~\ref{fig:strain}-(b) and (c),  the magnetic exchange interaction and Curie temperature  obtained by Monte Carlo simulation gradually decreases with strain. Curie temperature of 280 K can be achieved in YI$_2$ with a compression strain of -5\%. Even with tensile strain of 5\%, Curie temperature of the YI$_2$ is still larger than 160 K. Considered the enhanced Curie temperature and induced transition of VBM between $\Gamma$ and K/K\_, compression strain is thus suggested as a efficient strategy  to tune the performance of valleytronics of YX$_2$.

\section{\label{conclusion}CONCLUSION}
In summary, we have determined that single-layer 2H-YX$_2$(X=I, Br, Cl) are a series of stable two-dimensional ferrovalley semiconductors with a pair of energy valleys at K and K\_. Valley splitting occurs in YX$_2$ spontaneously, due to intrinsic magnetic exchange interactions and strong SOC. A large valley splitting of 108.98 meV was predicted in YI$_2$. The anomalous valley Hall effect are then proposed in YX$_2$ under in-plane electric field. In addition, compression strain is found to increase Curie temperature and  improve the performance of YX$_2$ in valleytronics. Our work provides a series of promising candidates for the realization of spontaneous valley polarization and controllable valley electronic devices, which is also expected to be verified experimentally in near future.
\begin{acknowledgments}
This work was supported by National Natural Science Foundation of China (Grant No. 11874092), the Fok Ying-Tong Education Foundation, China (Grant No. 161005), the Science Fund for Distinguished Young Scholars of Hunan Province(Grant No. 2021JJ10039),  the Planned Science and Technology Project of Hunan Province (Grant No.2017RS3034), Hunan Provincial Natural Science Foundation of China (Grant No.  2019JJ50636), Scientific Research Fund of Hunan Provincial Education Department (Grant No. 18C0227) and Open Research Fund of Key Laboratory of Low-Dimensional Quantum Structures and Quantum Control of Ministry of Education (Grant No. QSQC1902).
\end{acknowledgments}

%

\appendix*
\section{Band structure of YX$_2$(X=Br,Cl)}
\setcounter{figure}{0}
\renewcommand{\thefigure}{A\arabic{figure}}

\begin{figure*}
\centering
\includegraphics[width=0.65\textwidth]{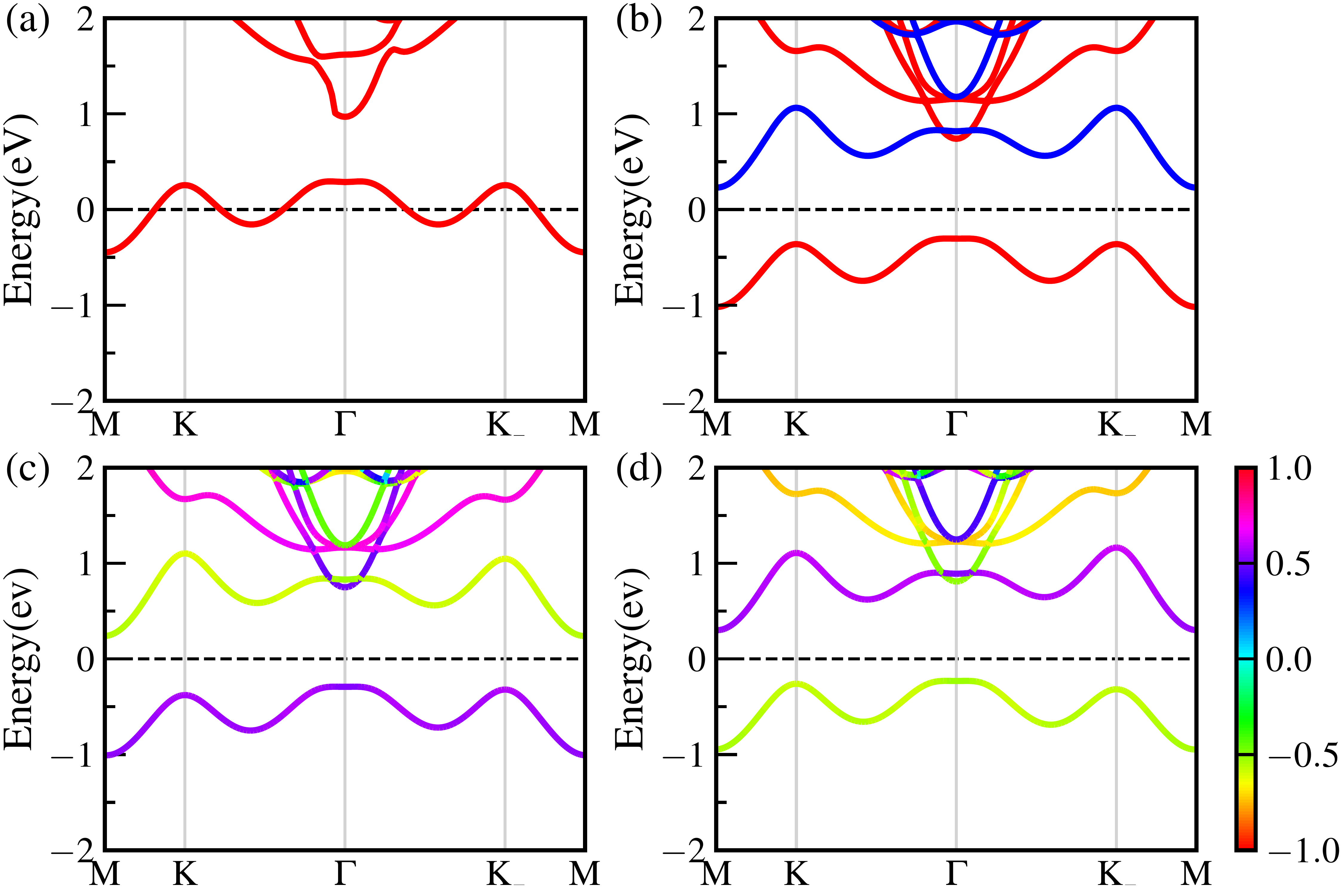}
\caption{\label{fig:Brband}(Color online)The band structure of SL YBr$_2$. (a) without spin-polarization and SOC. (b) with spin-polarization but without SOC.Red and blue lines respresent the spin-up and  spin-down bands. (c) both with spin-polarization and  SOC. (d) is the same as (c) but with the opposite magnetization orientation. The Fermi level is set to 0 eV. }
\end{figure*}

\begin{figure*}
\centering
\includegraphics[width=0.65\textwidth]{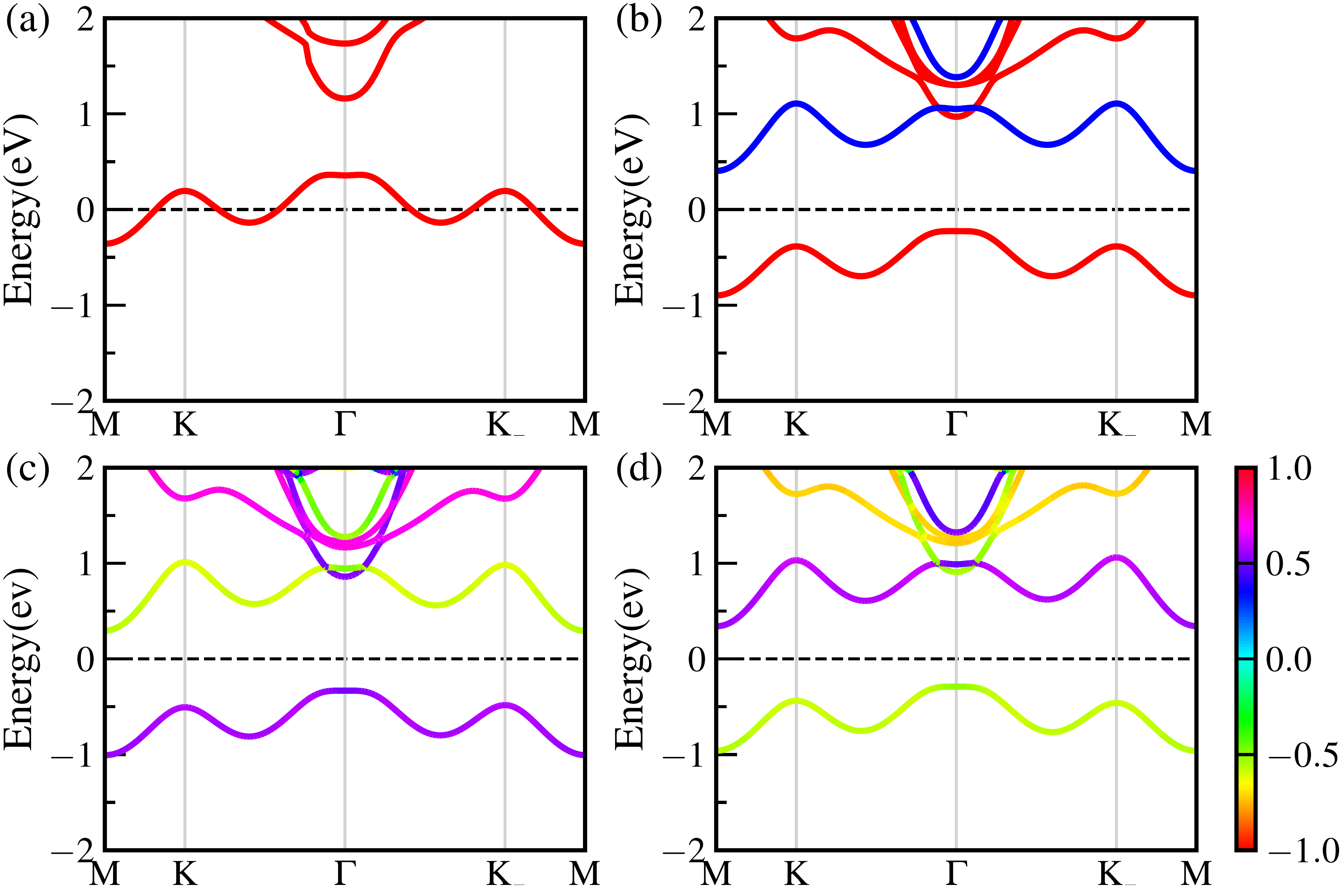}
\caption{\label{fig:Clband}(Color online)The band structure of SL YCl$_2$. (a) without spin-polarization and SOC. (b) with spin-polarization but without SOC.Red and blue lines respresent the spin-up and  spin-down bands. (c) both with spin-polarization and  SOC. (d) is the same as (c) but with the opposite magnetization orientation. The Fermi level is set to 0 eV. }
\end{figure*}

\begin{figure*}
\centering
\includegraphics[width=0.65\textwidth]{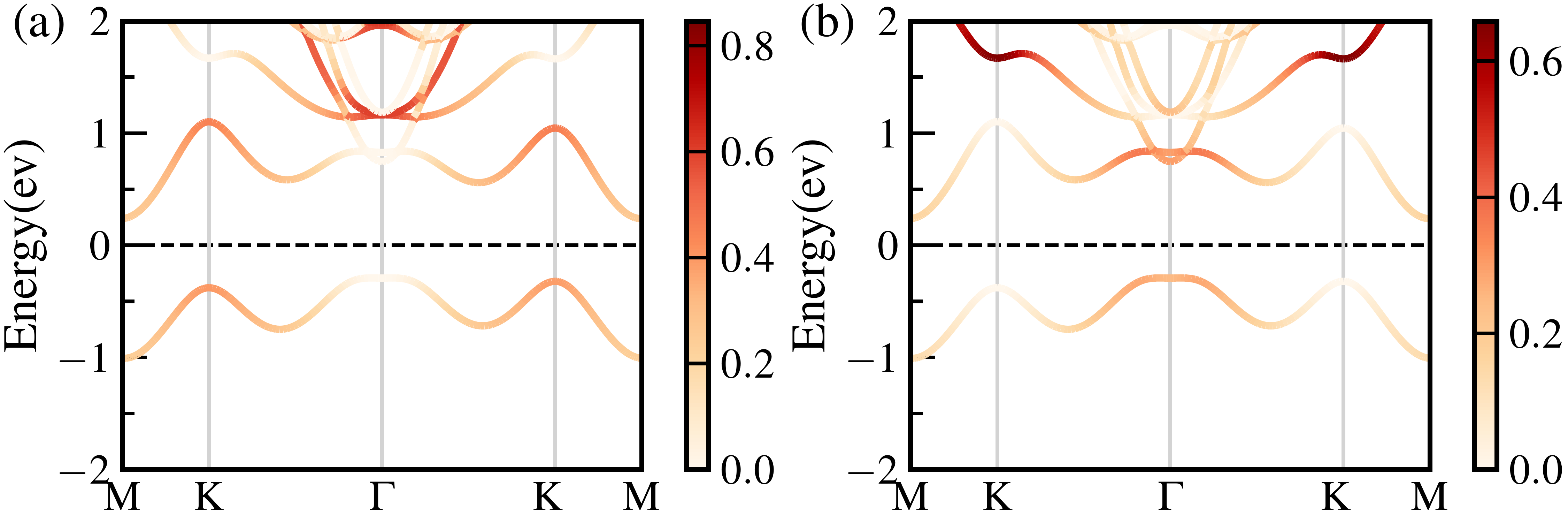}
\caption{\label{fig:Brfat_band}(Color online) The 4d orbital-resolved band structure of SL YBr$_2$. (a) and (b) represent d$_{xy}$(d$_{x^2-y^2}$) and  d$_{z^2}$, respectively.}
\end{figure*}

\begin{figure*}
\centering
\includegraphics[width=0.65\textwidth]{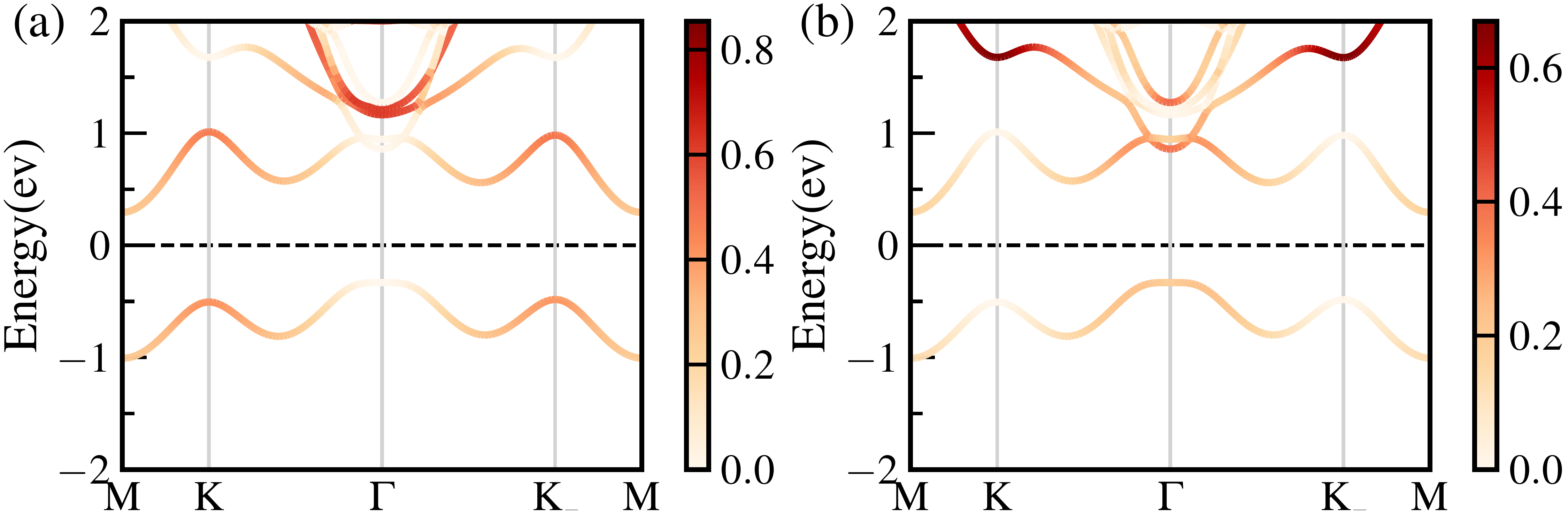}
\caption{\label{fig:Clfat_band}(Color online) The 4d orbital-resolved band structure of SL YCl$_2$. (a) and (b) represent d$_{xy}$(d$_{x^2-y^2}$) and  d$_{z^2}$, respectively.}
\end{figure*}
\end{document}